\documentclass[%
preprint,
superscriptaddress,
 amsmath,amssymb,
 aps,
prl,
]{revtex4-1}

\usepackage{graphicx}
\usepackage{dcolumn}
\usepackage{bm}
\usepackage{textcomp,gensymb} 
\usepackage {graphicx} 
\usepackage [caption = false] {subfig}

\usepackage{amssymb}

\usepackage{threeparttable}
\usepackage{multirow}
\usepackage{mathrsfs}
\DeclareMathAlphabet{\mathscrbf}{OMS}{mdugm}{b}{n}
\usepackage[usenames,dvipsnames]{color}
\usepackage{mathtools}

\newcommand{\vn}[1]{{\boldsymbol{#1}}}

\newcommand{\bege}{\begin{equation}}
\newcommand{\ee}{\end{equation}}
\newcommand{\bal}{\begin{aligned}}

\newcommand{\eal}{\end{aligned}}

\begin{document}

\title{Modification of Dzyaloshinskii-Moriya interaction stabilized domain wall chirality by driving currents}

\author{G.V. Karnad}
\affiliation{Institut f{\"u}r Physik, Johannes Gutenberg-Universit{\"a}t, Staudinger Weg 7, 55128 Mainz, Germany}

\author{F. Freimuth}
\affiliation{Peter  Gr{\"u}nberg  Institut  and  Institute  for  Advanced  Simulation,
Forschungszentrum  J{\"u}lich  and  JARA,  52425  J{\"u}lich,  Germany}

\author{E. Martinez}
\affiliation
{Departamento Fisica Applicada, Universidad de Salamanca, E37008 Salamanca,
Spain}

\author{R. Lo Conte }
\affiliation{Institut f{\"u}r Physik, Johannes Gutenberg-Universit{\"a}t, Staudinger Weg 7, 55128 Mainz, Germany}
\affiliation{Graduate School of Excellence \textquotedblleft{Materials Science in Mainz}\textquotedblright (MAINZ), Staudinger Weg 9, 55128 Mainz, Germany}

\author{G. Gubbiotti}
\affiliation
{Istituto Officina dei Materiali del CNR (CNR-IOM), Sede Secondaria di Perugia,
c/o Dipartimento di Fisica e Geologia, Universit{\'a} di Perugia, I-06123  Perugia, Italy}

\author{T. Schulz}
\affiliation{Institut f{\"u}r Physik, Johannes Gutenberg-Universit{\"a}t, Staudinger Weg 7, 55128 Mainz, Germany}

\author{S. Senz}
\affiliation{Max-Planck-Institut f{\"u}r Mikrostrukturphysik, 06120 Halle(Saale), Germany}

\author{B. Ocker }
\affiliation{Singulus Technology AG, 63796 Kahl am Main, Germany}

\author{Y. Mokrousov}
\email{y.mokrousov@fz-juelich.de}
\affiliation{Institut f{\"u}r Physik, Johannes Gutenberg-Universit{\"a}t, Staudinger Weg 7, 55128 Mainz, Germany}
\affiliation{Peter  Gr{\"u}nberg  Institut  and  Institute  for  Advanced  Simulation,
Forschungszentrum  J{\"u}lich  and  JARA,  52425  J{\"u}lich,  Germany}
\affiliation{Graduate School of Excellence \textquotedblleft{Materials Science in Mainz}\textquotedblright (MAINZ), Staudinger Weg 9, 55128 Mainz, Germany}

\author{M. Kl{\"a}ui}
\email{klaeui@uni-mainz.de}
\affiliation{Institut f{\"u}r Physik, Johannes Gutenberg-Universit{\"a}t, Staudinger Weg 7, 55128 Mainz, Germany}

\affiliation{Graduate School of Excellence \textquotedblleft{Materials Science in Mainz}\textquotedblright (MAINZ), Staudinger Weg 9, 55128 Mainz, Germany}

\date{\today}

\begin{abstract}

We measure and analyze the chirality of the Dzyaloshinskii-Moriya interaction (DMI) stabilized spin textures in multilayers of Ta$\mid$Co$_{20}$F$_{60}$B$_{20}$$\mid$MgO. The effective DMI is measured experimentally using domain wall motion measurements, both in the presence (using spin orbit torques) and absence of driving currents (using magnetic fields). 
We observe that the current-induced domain wall motion yields a change in effective DMI magnitude and opposite domain wall chirality when compared to field-induced domain wall motion (without current). We explore this effect, which we refer to as {\it current-induced DMI}, by providing possible explanations for its emergence, and explore the possibilty of
its manifestation in the framework of recent theoretical predictions of DMI  modifications due to spin currents.

\end{abstract}

\maketitle

The Dzyaloshinskii-Moriya interaction (DMI) \cite{DZYALOSHINSKY1958241,Moriya,fert1990} is an asymmetric exchange interaction which contributes to the stabilization of exotic spin textures such as chiral domain walls, spin spirals and skyrmions \cite{soumyanarayanan_emergent_2016}. DMI is found in systems possessing inversion asymmetry in bulk  as well as in multi-layers. The interfacial nature of the DMI in the latter case makes it possible to tailor the magnitude of the DMI by changing materials \cite{chen2013tailoring,torrejon_interface_2014}, layer ordering \cite{hrabec2014measuring}, ferromagnetic layer thickness \cite{LoConte_2017} and by interface modification \cite{hrabec2014measuring}.  Experimentally, there are several methods by which DMI has been measured and they are based on imaging domain wall (DWs) spin structures  \cite{chen2013tailoring, tetienne2015nature}, DW motion \cite{je_asymmetric_2013,emori2013current,ryu2013chiral,LoConte2015} or non-reciprocal spin wave dispersion measurements \cite{cho_thickness_2015,Belmeguenai_2015,Lee2016_BLSElectrical,Tacchi_2017}.

 Recently, a very intuitive relationship between the DMI and the ground state spin current has been found \cite{kikuchi2016dzyaloshinskii,freimuth2017relation}. The
ground state spin current represents the spin current present in an equilibrium system, in the absence of a net electric field. 
It was shown that linear contribution of the spin-orbit interaction (SOI) to the ground-state spin current is dominated by the Zeeman interaction of the spin-orbit field with the misalignment of the spins, which the conduction electrons acquire as they propagate in the spin textures \cite{freimuth2017relation}, resulting in an observation that to first order in spin-orbit the DMI is given by the ground state spin currents.
This finding directly suggests the possibility to tailor DMI by exciting the non-equilibrium spin currents in the system, for 
example by applying an external electric field, $\vn{E}$. It was shown that the corresponding effect of the DMI modified by spin current 
could be realized in a system where the spin polarization ($\vn{\sigma}$) of the spin current is perpendicular to the magnetization ($\vn{m}$), which makes magnetic multilayers with perpendicular magnetic anisotropy (PMA) the ideal candidates for the experimental 
observation of this new phenomenon \cite{freimuth2017relation}. However, as was roughly estimated for Co/Pt bilayers \cite{freimuth2017relation}, even for large spin currents of the order of  $10^7$\,A/cm$^2$\, $\hbar/e$, the resulting change of the DMI is on the order of 0.05 meV per atom, which is smaller than the DMI of the system in equilibrium by 2 orders of magnitude.
This implies that to observe this effect large spin-current densities and/or small values of the DMI in equilibrium are
required.

With this work we aim at exploring the effect that an electrical current can have on the DMI. Recent studies \cite{chenattukuzhiyil:tel-01274057,soucaille_probing_2016,LoConte_2017} have reported differences in magnitude and/or sign when comparing DMI extracted by techniques with and without the use of current. To study this in detail, we focus on a well characterized system, where we perform domain wall (DW) motion measurements: magnetic field-driven and current-driven, and probe the influence of the driving current on the effective measured DMI induced DW chirality. DW motion based techniques provide the possibility to perform experiments either by driving the domain walls with magnetic field or with currents, thus allowing for a direct comparison when using the same spin structures. In order to distinctly observe the influence of the current, we choose Ta (5) $\mid$Co$_{20}$F$_{60}$B$_{20}$ (0.8) $\mid$MgO (2) (all thicknesses in nm). 
This multilayer is a well studied system, and has a large perpendicular anisotropy \cite{ikeda2010perpendicular}, relatively small DMI \cite{torrejon_interface_2014,LoConte2015,tetienne2015nature,Karnad_Annihilation_arXiv2018} and a sizable spin Hall current source (Ta) \cite{liu_spin-torque_2012}. This makes it a suitable candidate to observe the influence of an electrical current on the DMI.

Field induced domain wall motion (FIDWM) in PMA materials relies on the application of a perpendicular field ($\vn{B_z}$) to modify the domain energy and thus cause domain wall motion. The experiments \cite{je_asymmetric_2013,hrabec2014measuring,kim_wide-range_2017} proposed to quantify the DMI using field driven expansion were mainly motivated by the ability to modify the domain wall energy density ($\sigma$) under application of an in-plane magnetic field ($\vn{B_x}$). Under simultaneous application of both an out-of-plane ($\vn{B_z}$, to drive the domain wall motion) and an in-plane ($\vn{B_x}$ to asymmetrically modify $\sigma$) magnetic field, it is possible to either slow down or speed up the motion of the domain wall \cite{je_asymmetric_2013,kim_wide-range_2017}. The sign of the in-plane magnetic field exploits the chirality induced in the domain walls by the DMI and thus selectively increases or decreases the domain wall energy, which directly affects the field ($\mu_0H_z$) driven wall velocity. This field-driven technique was mainly motivated \cite{hrabec2014measuring} by the principle of eliminating the influence of spin orbit torques in the measurement \cite{torrejon_interface_2014,LoConte2015} of DMI. However, it was recently shown \cite{vavnatka2015velocity,jue2016chiral} that this technique is not universally applicable and is dependent on the specifics of the material system, interfaces and motion regime.  
Nevertheless, they observed that this could be overcome by measuring systems in the flow regime, where pinning would not dominate the domain wall dynamics. This anomalous behavior can also be eliminated by using a system with low pinning \cite{burrowes2013low}.  Ju{\'e} et al., observed that in addition to the DMI induced chiral energy term there is an energy dissipation term that is also chiral \cite{jue2016chiral}. This was realized by measuring the domain wall dynamics in a system such as Pt$\mid$Co$\mid$Pt where the symmetry is weakly broken. This resulted in a negligible DMI and thus allowed them to observe experimental signals which could only be explained by the presence of a chiral damping term \cite{jue2016chiral}.

\begin{figure} 
     \includegraphics[width= 0.3\textwidth]{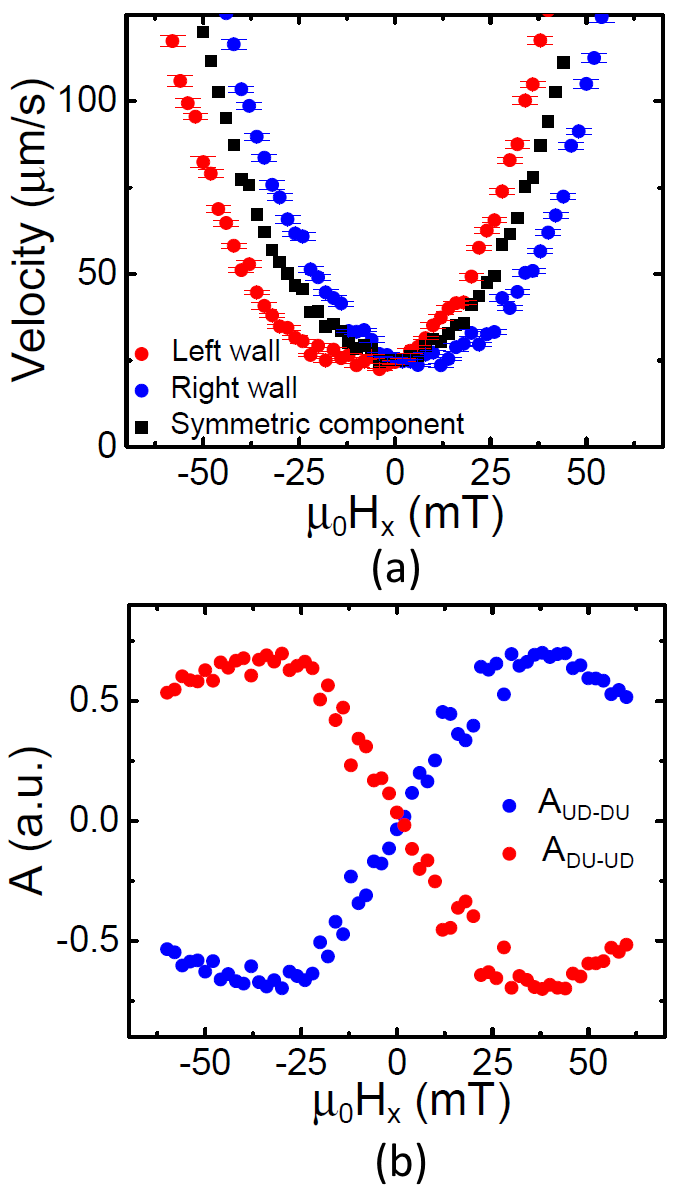}
     \caption{Field-induced domain wall motion experiment. a) DW velocity as a function of in-plane magnetic field ($\mu_0H_{x}$) where the DW is driven by an out-of-plane magnetic field ($\mu_0H_{z}$) of 1.5 mT. The symmetric component (black squares) is given by, S = (v$_{UD}$+v$_{DU}$)/2. b) The antisymmetric (blue dots - A$_{UD-DU}$ and red dots - A$_{DU-UD}$) component  of the DW velocity, where antisymmetric component is given by, A$_{UD-DU}$ = 2(v$_{UD}$-v$_{DU}$)/(v$_{UD}$+v$_{DU}$). 
     }
     \label{fig:DW_Bubble_experiment}
   \end{figure}

Here, we perform the asymmetric bubble expansion measurement (Fig. \ref{fig:DW_Bubble_experiment}) and observe that the application of an in-plane field indeed breaks the symmetry of the domain wall energy. We observe an asymmetric expansion of a bubble shaped domain \cite{Supplementary}. This indicates the presence of a chiral contribution dictating the domain wall dynamics: chiral energy (DMI) or chiral dissipation (chiral damping). We explore these possibilities, and separate the chiral effects \cite{jue2016chiral} by decomposing the domain wall velocities measured with respect to $\mu_0H_x$ into: a) Symmetric component: S = (v$_{UD}$+v$_{DU}$)/2 and b) Anti-symmetric component: A$_{UD-DU}$ = 2(v$_{UD}$-v$_{DU}$)/(v$_{UD}$+v$_{DU}$), as plotted in Fig. \ref{fig:DW_Bubble_experiment}. The anti-symmetric component clearly confirms the presence of a chiral term (either in energy or dissipation) in the system. We perform analytical calculations \cite{Supplementary}  to check the behavior of the domain wall velocities in the presence of only a chiral energy (DMI) in the system and observe that the numerical calculations reproduce the experimental observations. If the anti-symmetry (A) in the system was a result of the chiral damping, this would result in domain wall velocity curves which would not be possible to overlap despite translation along the x-axis \cite{jue2016chiral}. However, we observe that the domain wall velocities are symmetric around $\vert\mu_0H_{DMI}\vert$ and overlap by translating along the x-axis. This indicates that the dominant chiral effect in the system is the DMI, which is at the origin of the chiral energy in the system. The extracted $\mu_0H_{DMI}$ = 8.8 $\pm 2$ mT, the average of the in-plane field at which the DW velocity of left and right DW are the lowest (Bloch state) \cite{je_asymmetric_2013,hrabec2014measuring}. The effective DMI can be calculated by, $D_{eff} = \mu_0H_{DMI}M_s\Delta$, where saturation magnetization, $M_s$ = 0.705 $\times$ 10$^6$ A/m and the domain wall width, $\Delta$ =  5.35 nm.  This allows us to extract an effective DMI in the system, $D_{eff,field}$ = +33 $\pm 7.5$ $\mu$J/m$^{2}$. The symmetry of the asymmetric expansion also indicates that the system is right-handed.

To evaluate the influence of current on domain walls and hence on the DMI, we perform current induced domain wall motion (CIDWM) under the application of in-plane magnetic fields (details in the supplementary material \cite{Supplementary}). This method \cite{LoConte2015,torrejon_interface_2014} allows us to observe the influence of the spin orbit torques on the domain wall texture. The direction of the current-driven domain wall motion due to the spin Hall effect depends on the DW spin structure \cite{emori2013current,khvalkovskiy_matching_2013,ryu2013chiral}. This can be seen by the dependence of the damping-like (DL) torque on the magnetization direction \cite{emori2013current}. 
The direction of the CIDWM is governed by two important parameters: spin Hall angle and the chirality of the N\'eel wall induced by the DMI. The sign of the spin Hall angle is an intrinsic property \cite{tanaka_intrinsic_2008} of the heavy metal. We find from spin orbit torque measurements \cite{Schulz2017} that the sign of spin Hall angle of  Tantalum in our system is negative, which is in agreement with other theoretical \cite{tanaka_intrinsic_2008} and experimental \cite{liu_spin-torque_2012} results. Furthermore, from the field induced domain wall motion reported above, we know that the DMI is of relatively small magnitude (see also BLS measurements\cite{Supplementary}) and is right handed (see previous, D = +33 $\mu$J/m$^{2}$) in our system. Based on this, we expect the DWs to move in the direction of charge current.

\begin{figure} 
   
       \includegraphics[width= 0.3\textwidth]{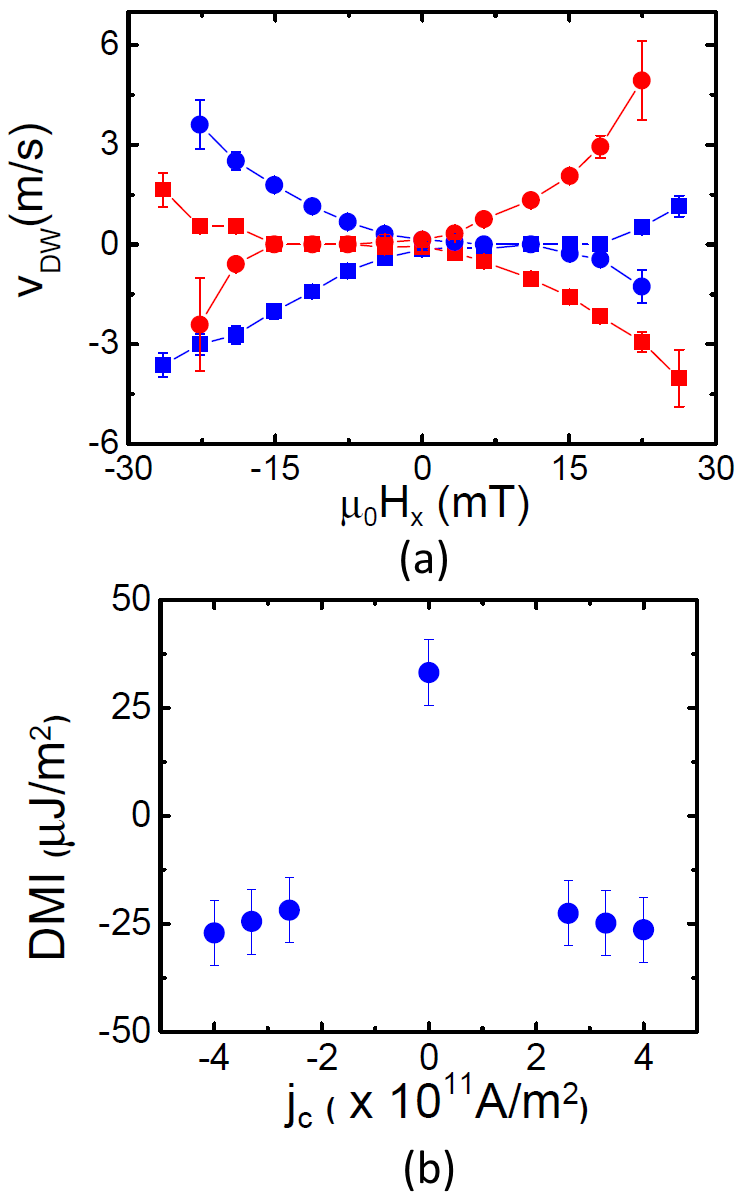}
     
     \caption{Current induced domain wall motion experiment a) DW velocities (Up-Down:Blue and Down-Up:Red) are measured under the application of an in-plane magnetic field. It is measured for a current density of + 3.3 $\times$ 10$^{11}$ A/m$^2$ (square) and - 3.3 $\times$ 10$^{11}$ A/m$^2$ (dots) b) The effective DMI for different current densities. A switching of DMI chirality is observed under injection of driving current.}
     \label{fig:CIDWM_chiral}
   \end{figure}

We perform the current induced domain wall motion experiments on an array of nanowires  (NWs) patterned on a thin film sample \cite{Supplementary}. The minimum current density to perform the experiment is dictated by the depinning current and the maximum current is limited by the thermal nucleation events. We observe that the domain walls move along the direction of electron flow which is opposite to the predicted direction. We observe that the velocity also increases on increasing the current density. To check for the presence and quantify the DMI, we perform the measurements for a range of current densities which provides stable domain wall motion without nucleation events (also in the presence of $\mu_0H_x$). We observe that the domain wall motion is indeed sensitive to the direction of the in-plane field (see Fig. \ref{fig:CIDWM_chiral} (a)). The domain walls stop moving for a certain in-plane field and then switch the direction of motion when the applied $\mu_0H_x$ is high. This stopping magnetic field can be interpreted as the effective DMI field, where the wall has a Bloch character and thus does not move as the effective spin orbit torque is zero, allowing us to extract an effective DMI field and a resultant DMI, $D$ \cite{torrejon_interface_2014}. This confirms that there is indeed DMI present in the system, and the motion of the DWs in the system is due to spin orbit torques. The DW motion direction however, can only be explained by an opposite chirality (Eqn. S2-S3 \cite{Supplementary}) compared to FIDWM. This indicates that the DMI in the system under the influence of current is switched to a left-handed system and reaches a value of $D_{eff,current}$ = - 26  $\pm 7.5 $ $\mu$J/m$^{2}$ at 4 $\times$ 10$^{11}$ A/m$^2$.

\begin{figure}
\includegraphics[width=1\linewidth]{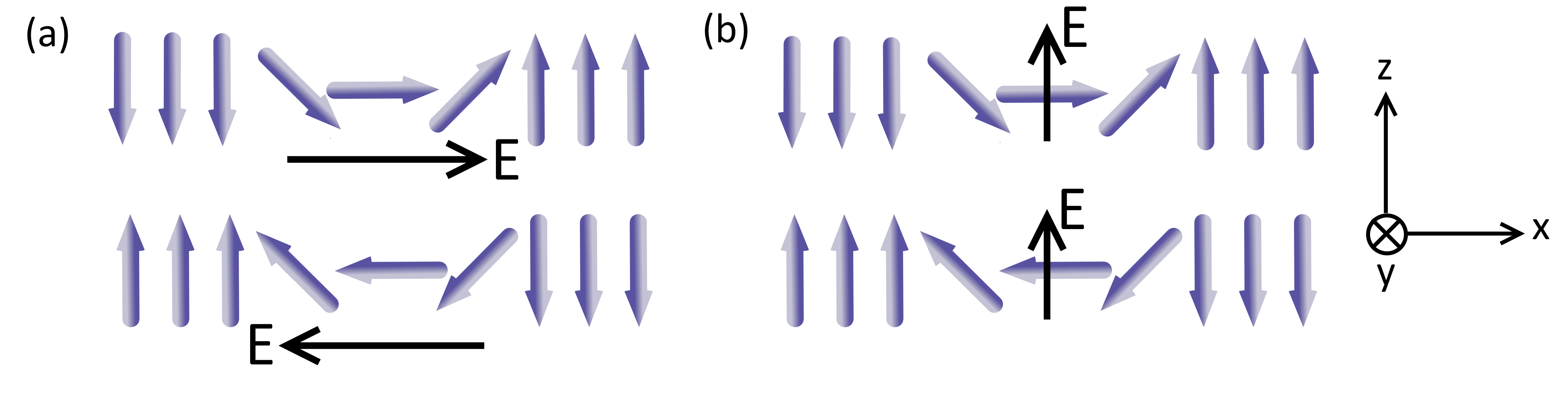}
\caption{\label{fig_neel_walls}
(a) A left-handed N\'eel-type down-up DW in the presence of
an electric field pointing to the right is symmetry-equivalent
to a left-handed N\'eel-type up-down DW in the presence of
an electric field pointing to the left. (b) A left-handed N\'eel-type down-up DW in the presence of
an electric field in $z$ direction is symmetry-equivalent
to a left-handed N\'eel-type up-down DW in the presence of
an electric field in $z$ direction.
}
\end{figure}

The results of the effective DMI as a function of current density (see Fig. \ref{fig:CIDWM_chiral} (b)) show a non-linear dependence, and we find in the experiment that the current-induced
change of DMI is manifestly independent of the polarity of the current (see Fig. \ref{fig:CIDWM_chiral} (b)). This is expected, because symmetry rules out a
current-induced modification
of DMI linear in the applied electric field in the magnetic bilayer 
geometry considered here.
To illustrate this we show in Fig.~\ref{fig_neel_walls}(a)
a chiral down-up N\'eel-type DW in the presence of an 
electric field $\vn{E}$ pointing to the right 
and a chiral up-down DW in the presence of an 
electric field $\vn{E}$ pointing to the left. Since a rotation 
around the interface normal by 180$^{\circ}$ maps the
two situations onto each other, the DW-width is not affected
to first order in $\vn{E}$. Consequently, there cannot be
a current-induced change of DMI linear in $\vn{E}$ in this
bilayer geometry.

The anomaly of sign difference \cite{chenattukuzhiyil:tel-01274057,soucaille_probing_2016}  and difference in magnitude \cite{LoConte_2017} in DMI has been mentioned in literature before. Fundamentally, the variation of the sign of DMI can be expected to arise from various origins. In addition to the current-induced modifications to the DMI brought by various spin currents which an electric field can induce in this complex
interfacial system, it is also expected that there is heating in the wires (increase of $\approx$ 50 K for the maximum current density applied here) caused by the current. This can increase the temperature \cite{torrejon_interface_2014} and can change the DMI \cite{kim2017microscopic}.

Other possible reasons for the DMI change include the change in sign of the damping-like torque, effects of the field-like torque on DW motion which were not taken into account, Oersted field, chiral damping, and, importantly the effects of the  spin transfer torque (STT) on the DW motion. We briefly discuss some of these possibilities: \textit{Chiral damping.}
Chiral damping and DMI have been proposed to share the same origin. The dominance of the chiral dissipation has been observed in a structure with minimal symmetry breaking \cite{jue2016chiral}. However, the system of Ta$\mid$Co$_{20}$F$_{60}$B$_{20}$$\mid$MgO has a large structural asymmetry and therefore the magnitude of the chiral damping can be expected to be minimal. Despite the issue of the FIDWM being in the creep regime, it should be noted that the reported sign of the DMI itself is consistent \cite{vavnatka2015velocity, kim2016universality} in the creep and flow regimes. \textit{Oersted field.}
Oersted field can be expected to influence the domain wall motion. This is especially true if the symmetry of field created along the NWs would be along the z-axis. However, while we calculate that the Oersted field is negligible, we eliminate its influence by measuring in an array of NWs - the neighboring NWs will largely compensate for any Oersted field in the structure. The influence of Oersted fields from the injection pads is also avoided by measuring at the central area of the NWs where Oersted field if any, would be zero. It should be noted that we do not observe any substantial difference in DW velocities in any of the NWs or along their individual length. \textit{Spin-transfer torque.}
During the injection of current pulses, due to the metallic nature of the stack there is always current shunting through the ferromagnetic layer as well. Due to the direction of current induced domain wall motion being with the electron flow direction, it would assist the effect of the SOT.  While the effect of the SHE-DL torque is sensitive to the domain wall structure, the adiabatic STT is not. Irrespective of the chirality or structure of the DW the motion would always be in the direction of the electron flow. However, we observe that the domain wall motion measured in our stacks is highly sensitive to the chirality of the domain wall and stops moving when the wall is tuned to the Bloch state. Additionally the current density in the FM is expected to be small and the irrelevance of STT for CIDWM in such thin multilayer stacks was also previously reported \cite{emori2013current}. In addition, due to the large thickness of Ta (5 nm), we estimate the spin-current density through CoFeB to be negligible ($<<$ 14\% of the total current density), suggesting a negligible STT. However, in the extreme case of an enhanced STT the direction of wall motion can flip sign, since the in-plane field changes the DW width and thus the non-adiabatic spin transfer torque also can potentially change sign \cite{Je_NegativeSTT_2017} and move the wall in the opposite direction than the adiabatic STT. However it should be noted that the scenario presented by Je \cite{Je_NegativeSTT_2017} et al., has a SHE compensated structure resulting in STT being the main driving force, unlike our case of a large SHE.

Finally, we propose examples for geometries in which symmetry allows for a
current-induced
modification of DMI linear in $\vn{E}$ according to the model put forward by \textcite{freimuth2017relation}. Potentially, this could lead to a more pronounced effect
as compared to the effect that we observe in this work.  As a first example,
we illustrate in Fig.~\ref{fig_neel_walls}(b) the case
in which the electric field is perpendicular
to the wall. Since rotation by 180$^{\circ}$ around the interface normal does not affect
the electric field vector in this case, a current-induced 
modification of DMI that is linear in $\vn{E}$ is allowed in
this case.

In order to obtain a systematic tool for the prediction of
current-induced DMI we introduce the DMI coefficients $D_{ij}$
such that
\bege
\delta F(\vn{r})=\sum_{ij}D_{ij}\hat{\vn{e}}_{i}\cdot
\left(
\hat{\vn{M}}
\times
\frac{\partial \hat{\vn{M}}}{\partial r_{j}}
\right)
\ee
is the change of energy-density due to DMI, 
where $\hat{\vn{M}}(\vn{r})$ is the magnetization direction
at position $\vn{r}$ and $\hat{\vn{e}}_{i}$ is a unit vector
pointing in the $i$-th Cartesian direction.
The DMI coefficients $D_{ij}$
have the symmetry properties of an axial tensor of
second rank~\cite{mothedmisot,itsot}. Similarly, spin currents are described by axial tensors of
second rank. Therefore,
when symmetry allows for a spin current $J_{j}^{i}$, 
where $j$ labels the direction of current flow and $i$ the orientation of the spins in the spin current, symmetry implies that the DMI coefficient $D_{ij}$ can be nonzero as well. If the electric field $\vn{E}$ induces a spin current $J_{j}^{i}$ in a magnetic system, we may therefore
expect that the DMI coefficient $D_{ij}$ changes as well.
This provides a systematic tool, for predicting cases in which 
DMI changes proportional to $\vn{E}$ can be observed.
This analogy between the spin current $J_{j}^{i}$ and the
DMI coefficient $D_{ij}$ follows from symmetry considerations alone,
because both quantities are axial tensors of second rank, and is therefore
generally valid. However,
it has been shown recently that at first order perturbation theory in
the spin-orbit interaction without applied electric field
even $J_{j}^{i}=-D_{ij}$ holds~\cite{freimuth2017relation,kikuchi2016dzyaloshinskii}, 
i.e., the DMI coefficient
is determined by the ground-state spin current. 
We can therefore understand why $\vn{E}$ can induce a change of DMI
in Fig.~\ref{fig_neel_walls}(b) from our knowledge of
the spin Hall effect: For the
N\'eel-type wall in Fig.~\ref{fig_neel_walls}(b) the wall-width depends on
$D_{yx}$ and an electric field in $z$ direction is expected to induce
a spin current $J_{x}^{y}$ via the spin Hall effect. This mechanism for the
current-induced DMI could be particularly relevant in the multi-layer
geometry, in which the magnetic layers are coupled by an antisymmetric 
RKKY-type of DMI ~\cite{Belmeguenai_2017_RKKY}.

Without applied electric field, $D_{ij}$ is even under time 
reversal~\cite{mothedmisot}.
Similarly, ground-state
spin currents are time-reversal-even.
The spin current generated by the spin Hall effect is
time-reversal-even as well.
In magnetic systems applied electric fields may also induce spin currents
that are time-reversal-odd. An example is the generation of a spin-current
by a polarizing magnet in a spin-valve setup, in which case the spin-current
changes sign when the magnetization of the polarizing magnet is reversed.
Such time-reversal-odd spin currents may induce changes of DMI as
well. Therefore, despite the absence of a complete overarching theory to understand our results, the experimental results indicate
that the DMI can indeed be influenced by the injection
of currents.

\begin{acknowledgments}
We acknowledge support by the EU
(Marie Curie ITN WALL -FP7-PEOPLE-2013-ITN 608031); the DFG (SFB TRR 173); Graduate School of Excellence Materials Science in Mainz (MAINZ) GSC 266. Y.M. and F.F. gratefully acknowledge computing time on the supercomputers of J\"ulich Supercomputing Center and RWTH Aachen University as well as funding by Deutsche Forschungsgemeinschaft (MO 1731/5-1).
\end{acknowledgments}


\end{document}